\documentclass{appolb}

\usepackage{graphicx}
\usepackage{hyperref}
\usepackage[numbers]{natbib}
\usepackage{amsfonts}

\providecommand\apj{ApJ}                 % {Ap. J.}
\providecommand\aap{A\&A}            % {A. \& A.} 
\providecommand\mnras{MNRAS}

\providecommand\cqg{CQG}
\providecommand\physrep{PhysRep}

\providecommand\nat{Nat.}

\newcommand\hGpc{\mbox{$h^{-1}$ Gpc}}
\newcommand\hMpc{\mbox{$h^{-1}$ Mpc}}

\newcommand\rC{R_{\mathrm{C}}}  %%% EDITOR modify as desired
\providecommand\gtapprox{\,\lower.6ex\hbox{$\buildrel >\over \sim$} \, }
\providecommand\ltapprox{\,\lower.6ex\hbox{$\buildrel <\over \sim$} \, }

\providecommand{\eprint}[1]{\href{http://arxiv.org/abs/#1}{{\tt [arXiv:#1]}}}
\providecommand{\url}[1]{\href{#1}{#1}}

\begin{document}
\title{Cosmic topology affects dynamics
%\thanks{{\ }}%
}
\author{Boudewijn F. Roukema
\address{Toru\'n Centre for Astronomy, Nicolaus Copernicus University,
ul. Gagarina 11, 87-100 Toru\'n, Poland}
%\and
%the Name(s) of other Author(s)
%\address{and their affiliation}
}
\maketitle
\begin{abstract}
%WCWC word count
The role of global topology in the dynamics of the Universe is
poorly understood. Along with observational programmes for
determining the topology of the Universe, some small theoretical
steps have recently been made. Heuristic Newtonian-like arguments 
suggest a topological acceleration effect that differs for differing
spatial sections. A relativistic spacetime solution solution shows
that the effect is not just a Newtonian artefact.
%WCWC word count
\end{abstract}
%PACS
%  98.80.Jk Mathematical and relativistic aspects of cosmology
%  04.20.Gz Spacetime topology, causal structure, spinor structure
%02.40.-k 	Geometry, differential geometry, and topology
%98.80.Es 	Observational cosmology (including Hubble constant, distance scale, cosmological constant, early Universe, etc) 
%maths numebrs
%85A40  %Astronomy and astrophysics / Cosmology
%83F05  %Relativity and gravitational theory  /  	Cosmology

\begin{figure}
  \centering 
    \includegraphics[width=0.5\textwidth]{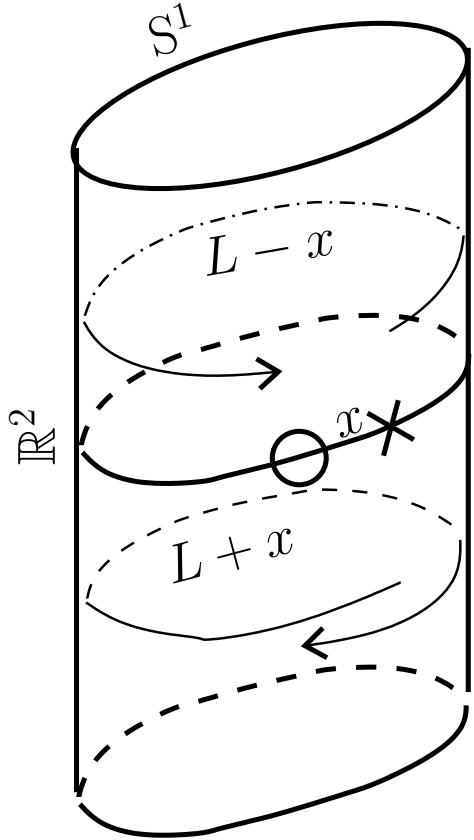} 
    \caption[]{T$^1$ spatial section, embedded and projected for
      convenience, showing how a negligible mass test particle (``X'')
      at distance $x$ from a massive particle (``O'') is subject to
      Newtonian accelerations from the two nearest topological images
      of the ``immediately'' nearby copy of the massive object
      (Sect.~\protect\ref{s-topacc-rel}).  See Fig.~1 of
      ref.~\protect\citep{ORB11} for the compact Schwarzschild-like,
      relativistic T$^1$ model (Sect.~\ref{s-topacc-rel}).
      \label{f-T1-accel}
    }
\end{figure} 
  
\section{{\em Rem quoque pr{\ae}cipuam, hoc est mundi formam}}

The shape of the Universe has long been considered to be an important
subject of study: {\em ``Rem quoque pr{\ae}cipuam, hoc est mundi
  formam, ac partium eius certam symmetriam non potuerunt invenire,
  vel ex illis colligere''} \citep[third page in preface, ref.~][]
{MK1543}. 
One of the aspects of ``shape'', the global geometrical
property {\em geometria situs} \citep{Euler1736} is now known as
``topology''.  Non-trivial topology of spatial sections of the
Universe has been discussed prior to \citep{Schw00}\footnote{English
  translation: ref.~\protect\citep{Schw98}.}  and since the beginnings of
relativistic cosmology \citep[e.g.,
][]{deSitt17,Fried23,Fried24,Lemaitre31ell}\footnote{Incomplete
  English translation, excluding observational analysis:
  ref.~\protect\citep{Lemaitre31elltrans}.}
\citep{Rob35}, along with the curvature of the spatial
sections. However, topology has generally been considered to be
unrelated to dynamics.  For example, Robertson \citep{Rob35} ``tacitly
assumed'' multiple connectedness for positively curved space while
commenting that ``we are still free to restore'' simple connectedness
and that the topology should be determined empirically. More recent
work has focussed on observational estimates of global toplogy where
observations are interpreted within the family of exactly homogeneous
Friedmann-Lema\^{\i}tre-Robertson-Walker (FLRW) models
\citep[e.g. ][and references
  therein]{LaLu95,Lum98,Stark98,LR99,BR99,RG04}.  One of the models
that has gained particular interest in order to match the cosmic
microwave background temperature fluctuations observed in the
Wilkinson Microwave Anisotropy Probe (WMAP) data is the Poincar\'e
\sloppy
dodecahedral space model
\citep{LumNat03,Aurich2005a,Aurich2005b,Gundermann2005,Caillerie07,RBSG08,RBG08,RK11}.

\section{Topological acceleration: Newtonian-like derivations}
 \label{s-topacc-newt}
Nevertheless, a Newtonian-like argument, motivated as a weak-field approximation of
would-be relativistic spacetime models, shows that the addition of a massive
particle to a homogeneous background implies an acceleration effect dependent
on the global topology of the spatial section \citep{RBBSJ06}. A 
negligible mass test particle displaced from the massive particle is 
subject to accelerations from images of the massive particle seen in 
approximately opposite directions in the covering space, with approximately
equal amplitudes (Fig.~\ref{f-T1-accel}).
These nearly cancel, but not quite. For a 
T$^1 := \mathbb{R}^2 \times \mathrm{S}^1$  spatial section, the resulting 
{\em topological acceleration} is linear to first order in $x/L$,
where $x$ is the test particle's displacement and $L$ is the length
of a closed spatial geodesic in the S$^1$ direction.

For 
a test particle displaced in a random direction from the
massive particle in
an exact T$^3$ spatial section, the linear effects 
from the topological images of the massive particle in many different
directions cancel \citep{RBBSJ06}, leaving a 
topological acceleration effect that is cubical in $x/L$
\citep{RR09}. The effect is again linear when a T$^3$ spatial
section has varying side lengths \citep{RBBSJ06}.

Similar Newtonian-like calculations in positively curved spaces are
less trivial. For the linear effect to cancel, the fundamental domain
presumably needs to have several closed spatial geodesics of the same length
and in different directions, in such a way that they can cancel the main
components of each other's topological accelerations.
There are three well-proportioned, multiply connected, 
positively curved spaces:
the octahedral space S$^3/\mathrm{T}^*$, the truncated cube space
S$^3/\mathrm{O}^*$, and the Poincar\'e dodecahedral space favoured
observationally, S$^3/\mathrm{I}^*$. The linear component of the topological acceleration
effect again cancels in the 
octahedral space and the truncated cube space \citep{RR09}.
In the Poincar\'e space, not only does the linear component cancel,
but the cubical component cancels too, leaving a topological acceleration
effect that is fifth order in $x/\rC$ ($\rC$ is the radius of curvature) \citep{RR09}.

Thus, the topological acceleration effect appears to mark
the Poincar\'e space---previously selected observationally as one
giving one of the best matches to the WMAP data---as being unique from
a theoretical point of view.

\section{Topological acceleration: relativistic} \label{s-topacc-rel}
Are the heuristic, Newtonian-like derivations of the topological acceleration effect 
valid relativistically? A first step in investigating this question 
is to study 
the compact Schwarzschild-like solution of the Einstein equations found
by Korotkin \& Nicolai \citep{KN94}. Outside of the event horizon, this spacetime
has T$^1$ spatial sections. 
Consider
a low-velocity test particle that is far $x \gg GM$ from the black hole's event horizon
$GM$ (in Weyl coordinates, not Schwarzschild coordinates; $G$ is the
gravitational constant),
in a model where the spatial geodesic length $L$ is also much greater than the
test particle's distance from the black hole centre, i.e. $ 0 < GM \ll x \ll L$.
If the heuristic, Newtonian-like derivation is correct,
then the test particle in this case
should be subject to a four-acceleration whose spatial component gives the
result found earlier. This is indeed the case \citep{ORB11}. 
Numerically, for low-velocity test particles, the linear expression
$4\zeta(3) GML^{-3}\,x$, where $\zeta(3)$ is Ap\'ery's constant,
is a good approximation,
to within $\pm10\%$, over $3{\hMpc} \ltapprox x \ltapprox  2{\hGpc}$,
if the massive particle is at a cluster scale, $M \sim 10^{14} M_\odot$,
and the closed spatial geodesic length is $L \sim 10$ to $20{\hGpc}$ \citep{ORB11}.

\section{Prospects}

It would be good to perform 
relativistic calculations of topological acceleration in expanding,
FLRW-like models of T$^3$, S$^3/\mathrm{T}^*$ S$^3/\mathrm{O}^*$, and
S$^3/\mathrm{I}^*$ spatial sections. This is not just needed
to compare with astronomical observations, but also for theoretical
work in early universe studies. This is needed independent of
spacetime theories that extend beyond four dimensions.
Exact solutions of the Einstein
equations 
for S$^3/\mathrm{T}^*$,  S$^3/\mathrm{O}^*$, and
S$^3/\mathrm{I}^*$ models containing a massive particle above
a homogeneous background seem unlikely to exist, 
so this would probably require numerical work. Numerical experience
with $N$-body simulations calculated using Newtonian gravity on
an FLRW background expanding universe model \citep[e.g. ][]{Dehnen11}
shows that these are not easy. Equivalent numerical calculations 
that are relativistic and  background-free
are unlikely to be easier, but may be unavoidable.

%{\em ``Rem quoque pr{\ae}cipuam, hoc est mundi foramm, ac partium eius certam
%symmetriam non potuerunt invenire, vel ex illis colligere''}

%google: Also the best of the thing, that is, shape of the world, ac a certain
%symmetry of its parts they could not find, and / or to collect one of
%them;

%% CUT_HERE

\end{document}